\documentclass[final]{aipproc}

\layoutstyle{8x11double}

\usepackage{epsfig}
\usepackage{graphicx}

\usepackage{amsfonts}
\usepackage{dcolumn}

\def\slashchar#1{\setbox0=\hbox{$#1$}
   \dimen0=\wd0 \setbox1=\hbox{/} \dimen1=\wd1
   \ifdim\dimen0>\dimen1 \rlap{\hbox to \dimen0{\hfil/\hfil}} #1
   \else  \rlap{\hbox to \dimen1{\hfil$#1$\hfil}} / \fi}

\begin{document}

\title{Scalar-isoscalar states, gravitational form factors, and dimension-2 condensates in a large-$N_c$ Regge approach}

\classification{12.38.Lg, 11.30, 12.38.-t}
\keywords      {{$\sigma$ meson, scalar-isoscalar states, large-$N_c$ Regge
  models, pion and nucleon gravitational form
  factors, dimension-2 condensate}}

\author{\underline{Enrique Ruiz Arriola}~\thanks{Speaker at 
9th Conference On Quark Confinement and the Hadron Spectrum (Confinement IX),
	30 Aug. - 3 Sept. 2010, Madrid, Spain}{~}}
{address={Departamento de
  F\'{\i}sica At\'omica, Molecular y Nuclear, Universidad de Granada, E-18071 Granada, Spain}}
\author{Wojciech Broniowski} 
{address={The H. Niewodnicza\'nski Institute of Nuclear Physics, Polish Academy of Sciences, PL-31342 Krak\'ow, Poland},
altaddress={Institute of Physics, Jan Kochanowski University, PL-25406~Kielce, Poland}}

\begin{abstract} 
Scalar-isoscalar states ($J^{PC}=0^{++}$) are analyzed within the
large-$N_c$ Regge approach. We find that the lightest $f_0(600)$
scalar-isoscalar state fits very well into the pattern of the radial
Regge trajectory. We confirm the obtained mass values from an analysis
of the pion and nucleon spin-0 gravitational form factors, recently
measured on the lattice.  We find that a simple two-state model
suggests a meson nature of $f_0(600)$, and a glueball nature of
$f_0(980)$, which naturally explains the ratios of various coupling
constants.  Finally, matching to the OPE requires a fine-tuned mass
condition of the vanishing dimension-2 condensate in the Regge
approach with infinitely many scalar-isoscalar states.
\end{abstract}

\maketitle

\section{Introduction \label{sec:intro}}

Hadron resonances appearing in the PDG tables increase their mass
while their width remains constant~\cite{Amsler:2008zz}. On the other
hand, in the large-$N_c$ limit, with $g^2 N_c$ fixed, mesons and
glueballs are stable, their masses become independent on $N_c$,
$m={\cal O}(N_c^0)$, while their widths $\Gamma$ are suppressed as
$1/N_c$ and $1/N_c^2$, respectively, which means that $\Gamma/m$ is
suppressed (see e.g. \cite{ Pich:2002xy} for a review).  This suggests
that excited states in the mesonic spectrum may follow a reliable
large-$N_c$ pattern. The main feature of a resonance is that it
corresponds to a mass distribution, with values approximately spanning
the $m \pm \Gamma/2$ mass interval.  The lowest resonance in QCD is
the $0^{++}$ state $f_0(600)$ or the $\sigma-$meson, which appears as
a complex pole in the second Riemann sheet of the $\pi\pi$ scattering
amplitude at $s_\sigma = m_\sigma^2 - i m_\sigma \Gamma_\sigma$ with
$m_\sigma=347(17) {\rm MeV}$ and $\Gamma_\sigma/2= 345(24) {\rm
MeV}$~\cite{Caprini:2005zr} (see also Ref.~\cite{Kaminski:2006qe}).
Higher $0^{++}$ states are listed in Table~\ref{tab:regge-fits}.

For scalar states a measure of the spectrum is given in terms of the
trace of the energy momentum tensor~\cite{Donoghue:1991qv}
\begin{eqnarray}
\Theta^\mu_\mu \equiv  \Theta  = 
\frac{\beta(\alpha)}{2\alpha}
G^{\mu\nu a} G_{\mu\nu}^a + \sum_q m_q \left[ 1 + \gamma_m (\alpha)
\right] \bar q q . 
\label{def:theta}
\end{eqnarray} 
Here $\beta(\alpha) = \mu^2 d \alpha / d\mu^2 $ denotes the beta
function, $\alpha=g^2/(4\pi)$ is the running coupling constant,
$\gamma_m(\alpha) = d \log m / d \log \mu^2 $ is the anomalous
dimension of the current quark mass $m_q$, and $G_{\mu\nu}^a$ is the
field strength tensor of the gluon field. This operator connects
scalar states to the vacuum through the matrix element
\begin{eqnarray}
 \langle 0 | \Theta | n
\rangle = m_n^2 f_n \label{dM}.
\end{eqnarray}
The two-point correlator reads 
\begin{eqnarray}
&& \Pi_{\Theta \Theta} (q) = i \int d^4 x e^{i q \cdot x} \langle 0 |
 T \left\{ \Theta (x) \Theta (0) \right\} | 0 \rangle
 \nonumber \\ &=& \sum_n \frac{f_n^2 q^4}{m_n^2 -q^2}
= q^4 \left[ C_0 \log q^2 + \sum_{n} \frac{C_{2n}}{q^{2n}} \right], \label{qcdsr}
\end{eqnarray} 
where in the second line we saturate with scalar states and the large
$q^2 \gg \Lambda_{\rm QCD}$ limit is taken. Comparison with the
Operator Product Expansion (OPE)~\cite{Narison:1996fm} leads to
\begin{eqnarray}
C_0 &=& 
-\lim_{n \to \infty }\,  \frac{f_n^2}{d  m_n^2 /dn }
=  - \frac{ N_c^2-1}{2\pi^2} \left(\frac{\beta(\alpha)}{\alpha}\right)^2
 \label{eq:C0},  \\ 
C_2 &=&  \sum_n f_n^2  \qquad \qquad = 0 \,   \label{eq:C2},  \\ 
C_4 &=&  \sum_n f_n^2 m_n^2 \qquad \quad = \left(\frac{\beta(\alpha)}{\alpha}\right)^2 \langle G^2 \rangle \,. 
\end{eqnarray} 
Equation~(\ref{eq:C0}) requires infinitely many states, while
Eq.~(\ref{eq:C2}) suggests a positive and non-vanishing 
gauge-invariant dimension-2 object,  $C_2= i \int d^4 x x^2 \langle \Theta
(x) \Theta \rangle $, which is generally
non-local, as it should not appear in the OPE.

\begin{figure}[tb]
\includegraphics[width=.47\textwidth]{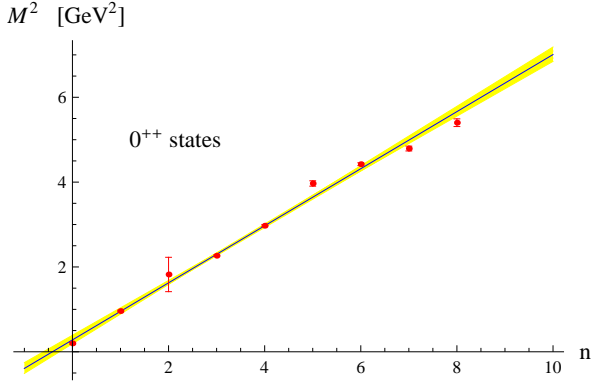}
\caption{Radial Regge trajectory corresponding to the squared mass of all
$J^{PC}=0^{++}$ scalar-isoscalar states listed in the
PDG tables~\cite{Amsler:2008zz}.  The four heaviest $0^{++}$ states are not
yet well established and are omitted from the PDG summary table. 
The error bars correspond to the errors in the determination of the 
square of mass, listed in~\cite{Amsler:2008zz}. The straight line is the result of our fit.}
\label{fig:regge-trajectory}
\end{figure}

\section{Scalar Regge spectrum}

Radial and rotational Regge trajectories were analyzed in
Ref.~\cite{Anisovich:2000kxa}. In Ref.~\cite{Anisovich:2005jn} the
scalar sector was studied in more detail. Two parallel radial
trajectories could then be identified, including three states per
trajectory. In a recent work~\cite{RuizArriola:2010fj} we have
analyzed all the $0^{++}$ states which appear in the PDG tables (see
Fig.~\ref{fig:regge-trajectory}) and found that {\it all} fit into a
single radial Regge trajectory of the form
\begin{eqnarray}
M_S (n)^2 = \frac{a}{2} n + m_\sigma^2. \label{eq:massS}
\end{eqnarray} 
The mass of the $\sigma$ state can be deduced from this trajectory as
the mass of the lowest state. The resonance nature of these states
suggests using the corresponding half-width as the mass uncertainty by
minimizing
\begin{eqnarray}
\chi^2 = \sum_n \left(\frac{M_{f,n}-M_S(n)}{\Gamma_{f,n}/2} \right)^2 ,
\label{eq:chi2}
\end{eqnarray}
which yields $\chi^2 /{\rm DOF} = 0.12 $ (see also
Table~\ref{tab:regge-fits}) with 
\begin{eqnarray}
a = 1.31(12) {\rm ~GeV}^2, \quad \; m_\sigma = 556(127) {\rm ~MeV} \,  \label{fitII}.  
\end{eqnarray} 
Formula (\ref{eq:massS}) is equivalent to two parallel radial Regge
trajectories with the {\em standard} slope
\begin{eqnarray}
M_{S,-} (n)^2 &=& a \, n + m_\sigma^2,  \label{traj-1}  \\
M_{S,+} (n)^2 &=& a \, n + m_\sigma^2 + \frac{a}{2}  , \label{traj-2}
\end{eqnarray} 
where $a=2 \pi \sigma$, and $\sigma$ is the string tension associated
to the potential $V(r)= \sigma r$ between heavy colored sources.
The value $\sqrt{\sigma} = 456(21) \, {\rm MeV}$ agrees well with lattice
determinations of $\sqrt{\sigma} = 420 {\rm MeV}$~\cite{Kaczmarek:2005ui}.

This situation suggests the existence of a hidden symmetry in the
$0^{++}$ sector. In the holographic approach based on the AdS/CFT
correspondence the symmetry corresponds to parity in the
fifth-dimensional variable. This is similar to the one-dimensional
harmonic oscillator; all states with the energy $E_n = \hbar \omega
(n+1/2)$ can be separated in parity {\it even} and {\it odd } states,
with energies $E_n^{(+)} = 2 \hbar \omega ( n+1/4)$ and $E_n^{(-)} = 2
\hbar \omega ( n+3/4)$, respectively, having twice the slope of $E_n$.

\begin{table}[tb]
\begin{tabular}{|c|c|c|c|c|}
\hline 
Resonance & $M$ [MeV] & $\Gamma$ [MeV] & $n$ & $M$ (Fit) \\
\hline 
$f_0(600)$    & $400-1200$      & $500-1000$    & 0 & $556$       \\ 
$f_0(980)$    & $980(10)$      & $70(30)$ & 1  & $983$    \\ 
$f_0(1370)$   & $1350(150) $      & $400(100)$  & 2  & $1274$        \\ 
$f_0(1500)$   & $1505(6) $      & $109(7)$   & 3 & $1510$      \\ 
$f_0(1710)$   & $1724(7) $      & $137(8)$   & 4  &$1714$     \\ 
$f_0(2020)$   & $1992(16) $      & $442(60)$ & 5  &$1896$       \\ 
$f_0(2100)$   & $2103(8) $      & $209(19)$  & 6  &$2062$      \\ 
$f_0(2200)$   & $2189(13) $      & $238(50)$ & 7  &$2215$       \\ 
$f_0(2330)$   & $2321(30) $      &  $223(30)$ & 8 & $2359$         \\ 
\hline
\end{tabular}
\caption{\label{tab:regge-fits} PDG values of resonance
  parameters~\cite{Amsler:2008zz}, compared to the fit to the radial
  Regge spectrum of the $0^{++}$ scalar-isoscalar states, $M_n^2
  = \frac12 a n + m_\sigma^2 $, given by Eq.~(\ref{eq:chi2}). The
  lowest $n=0$ state, corresponding to the $f_0(600)$ resonance, is
  excluded from the fit.}
\end{table}

Besides, there seems to be no obvious difference between
mesons and glueballs, as far as the spectrum is concerned. Note that the
Casimir scaling suggests that the string tension is $ \sigma_{\rm
glueball}= \frac94 \sigma_{\rm meson}$, but this holds in the case of fixed
and heavy sources. The fact that we have light quarks might explain why
we cannot allocate easily the Casimir scaling pattern in the light-quark scalar-isoscalar
spectrum.

\section{Gravitational form factors}

Hadronic matrix elements of the energy-momentum tensor, the so-called
gravitational form factors (GFF) of the pion and nucleon, correspond
to a dominance of scalar states in the large-$N_c$ picture, as ($u(p)$
is a Dirac spinor)
\begin{eqnarray}
\langle \pi (p') | \Theta | \pi(p) \rangle &=& 
\sum_n \frac{g_{n \pi \pi} f_n q^2
m_n^2}{m_n^2-q^2}, \\ 
\langle N(p') | \Theta | N(p) \rangle &=& \bar u(p') u(p) \sum_n  \frac{   g_{n NN} f_n  m_n^2}{m_n^2-q^2},
\end{eqnarray} 
where the sum rules $\sum_n {g_{n\pi \pi}}{f_n}
=1$~\cite{Narison:1988ts} $M_N = \sum_n g_{n NN}
f_n$~\cite{Carruthers:1971vz} hold. Unfortunately, the lattice QCD
data for the pion~\cite{Brommel:2007xd} and nucleon (LHPC
~\cite{Hagler:2007xi} and QCDSF~\cite{Gockeler:2003jfa}
collaborations), picking the valence quark contribution, are too noisy
as to pin down the coupling of the excited scalar-isoscalar states to
the energy-momentum tensor.  Nevertheless, useful information
confirming the (Regge) mass estimates for the $\sigma$-meson can be
extracted using multiplicative QCD evolution of the GFF through the
valence quark momentum fraction, $\langle x \rangle_{u+d}$, as seen in
deep inelastic scattering or on the lattice at the scale $\mu= 2{\rm
GeV}$. For the pion GFF we obtain the fit
\begin{eqnarray}
\langle x \rangle_{u+d}^\pi = 0.52(2), \qquad m_\sigma = 445(32) {\rm ~MeV} \, ,  \label{opti2}
\end{eqnarray}
whereas for the nucleon GFF we get
\begin{eqnarray}
\langle x \rangle_{u+d}^N = 0.447(14), \qquad \ m_\sigma = 550^{+180}_{-200} {\rm MeV}.
\end{eqnarray}
Assuming a simple dependence of $m_\sigma$ on $m_\pi$,
\begin{eqnarray}
m_\sigma^2(m_\pi)=m_\sigma^2+c\left  ( m_\pi^2-m_{\pi,{\rm phys}}^ 2 \right ), \label{msmpi}
\end{eqnarray}
yields $ m_\sigma = 550^{+180}_{-200} {\rm MeV}$ and $
c=0.95^{+0.80}_{-0.75}$, or $\ m_\sigma = 600^{+80}_{-100} {\rm MeV} $
and $ c=0.8(2)$, depending on the lattice data~\cite{Hagler:2007xi}
or~\cite{Gockeler:2003jfa}, respectively.   {\it
Higher} quark masses might possibly clarify whether or not the state
evolves into a glueball or a meson, since in that case one has,
respectively, either $m_\sigma /(2 m_q)\to 0 $, or $m_\sigma
/(2m_q) \to 1$.

\section{Dimension-2 condensates}

It is interesting to discern the nature of the $\sigma$ state from an
analysis of a truncated spectrum. The minimum number of states,
allowed by certain sum rules and low energy theorems, is just two. In
Ref.~\cite{RuizArriola:2010fj} we undertake such an analysis, which
suggests that $f_0(600)$ (denoted as $\sigma$) is a $\bar q q $ meson,
while $f_0(980)$ (denoted as $f_0)$ is a glueball according to the
$N_c$ scaling of various quantities (see Table~\ref{tab:nc}). The argument is based on the fact that one
obtains $f_\sigma/f_{f_0} \sim g_{f_0 NN}/g_{\sigma
NN} \sim g_{f_0 \pi\pi}/g_{\sigma \pi\pi} \sim 1/\sqrt{3} $, however $C_2 = 
f_\sigma^2+f_{f_0}^2 \neq
0$ because the number of states is finite.
\begin{table}[tb]
\caption{$N_c$-scaling. \label{tab:nc}}
\begin{tabular}{|c|c|c|} 
\hline
quantity & glueball & $q \bar q$ meson \\ \hline
$m_n$ & 1  & 1 \\
$f_n$    & $N_c$ & $\sqrt{N_c}$ \\
$\Gamma_{n\pi\pi}$ & $1/{N_c^2}$  & $1/N_c$ \\
$g_{n\pi\pi}$ & $1/N_c$  & $1/\sqrt{N_c}$ \\
$g_{nNN}$ & $1$  & $\sqrt{N_c}$ \\
\hline
\end{tabular}
\end{table}

The infinite Regge spectrum of Eq.~ (\ref{eq:massS}) with
 Eq.~(\ref{eq:C0}) may be modeled with a constant $f_{f_0}=f_{n,+}={\cal
 O}(N_c)$  discarding $f_\sigma = f_{n,-}={\cal O} (\sqrt{N_c})$
 . Naively, we get $C_2=\infty$. However, $C_2$ may vanish, as
 required by standard OPE, when infinitely many states are considered
 {\it after regularization}. Using $\zeta$-function regularization~\cite{Arriola:2006sv}, yields~\cite{RuizArriola:2010fj}
\begin{eqnarray}
C_2 \equiv  \lim_{s\to 0} \sum f_{n,+}^2 M_{S,+} (n)^{2s}   
 = f_{f_0}^2 \left( 1/2 - m_{f_0}^2/a \right)\, , \label{match:dim2}  
\end{eqnarray}
which at leading $N_c$ implies $C_2=0$ for $m_{f_0} = \sqrt{a/2}=
 810(40) {\rm MeV} $, a reasonable value to ${\cal O}(1/N_c)$.

What should these $m_\sigma$ values be compared to?  Besides the pole
definition one also has the Breit-Wigner (BW) definition, which in the
$\pi\pi$ data is disputed for the $\sigma$ but not for the $\rho$. Our
analysis is driven by large-$N_c$ considerations (see also
Refs.~\cite{Pelaez:2009eu,RuizdeElvira:2010cs}). In
Ref.~\cite{Nieves:2009kh} it was shown that the difference between the
BW and the pole definitions is ${\cal O}(1/N_c^2)$ and, further, that
for $N_c=3$ the BW definition works {\it as good} for the $\sigma$ as for the
$\rho$. In Ref.~\cite{Nieves:2009ez} it was argued that $m_\sigma -
m_\rho = {\cal O} (1/N_c)$.  Thus, we expect the present estimates to
be in between, incorporating a systematic ${\cal O}(1/N_c)$ mass
shift.

\begin{theacknowledgments}
We thank Ximo Prades for useful discussions. He had planned to attend
this meeting, but he died after a long illness which he faced squarely
with formidable endurance up to the last moment. We will miss him!!

This work is partially supported by the Polish Ministry of Science and
Higher Education (grants N~N202~263438 and N~N202~249235), Spanish
DGI and FEDER funds (grant FIS2008-01143/FIS), Junta de Andaluc{\'\i}a
(grant FQM225-05).
\end{theacknowledgments}


\bibliographystyle{aipproc}   

\end{document}